\def\beq{\begin{equation}}
\def\eeq{\end{equation}}
\def\bmulteq{\begin{eqnarray}}
\def\emulteq{\end{eqnarray}}
\def\>{\rangle}
\def\<{\langle}
\def\tqi{{\mathcal{T}}}

\documentclass[twocolumn,showpacs,preprintnumbers,amsmath,amssymb]{revtex4}

\begin{document}

\preprint{BNL-HET-02-26}
\title{What is the speed of quantum information?}
\author{Robert Garisto}
\affiliation{BNL Theory Group, Bldg 510a, Brookhaven National Lab, Upton, NY 11973, U.S.A.}

\begin{abstract}
We study the apparent nonlocality of quantum mechanics as a transport
problem.  If space is a physical entity through which quantum
information (QI) must be transported, then one can define its speed.
If not, QI exists apart from space, making space in some sense
`nonphysical'.  But we can still assign a `speed' of QI to such models
based on their properties.  In both cases, classical information must
still travel at $c$, though in the latter case the origin of local
spacetime itself is a puzzle.  We consider the properties of different
regimes for this speed of QI, and relevant quantum interpretations.
For example, we show that the Many Worlds Interpretation (MWI) is
nonlocal because it is what we call {\it spatially complete}.
\end{abstract}

\maketitle

Since the seminal work of Einstein, Podolsky and Rosen \cite{EPR}, it
has been clear that quantum mechanics appears to be nonlocal.  Of
course, classical observables respect relativistic causality, but this
was imposed upon the theory {\it a posteriori} by insisting that
commutators of spacelike operators vanish.  Bell showed that quantum
mechanics violates an inequality which is satisfied by any model with
local hidden variables
\cite{Bell}, and experiments agree with quantum predictions \cite{Aspect}.  
So there is something beyond classical information in quantum theory,
and if we were to try to model it in terms of hidden variables, it
would behave nonlocally.  We believe that this apparent nonlocality
must either be reified or explained away in any complete theory.

By quantum information (QI), we mean the set of all dynamical entities in
the formalism of a complete quantum theory.
QI could include `information' about the wave function, phases,
entanglement, histories, or something else.
Although QI {\it could} specify unique outcomes for observables, like
the nonlocal hidden variables of Bohm's model, and our results should
apply to such models, we will stress more the possibility that QI does
{\it not} involve hidden variables.  Instead, QI would allow a deeper
understanding of the theory without adding classical determinism.  
In any case, our goal is simply to categorize the spacetime behavior this
QI can have, and ask questions such as ``Is it local?'', ``Is it
causal?'' and even more simply, ``Where is it?''

In classical mechanics, information resides in physical space.  To
affect another location, it must be transported across space, with
limiting speed $c$---{\it i.e.}, it can  affect things only in its
forward lightcone.  For QI, one possibility is that space is
`physical' in this way, and QI exists in it.    
Then QI needs transport through space at some speed $v_{QI}$, and the theory
should explain the dynamics of this new kind of transport.
Alternatively, space could be `nonphysical' in that
QI does not exist in it and thus does not need transport through it.  
As we will see later, it is still useful to characterize this kind of
world by a `speed'.  So either way, we believe that any complete theory
should give a definitive answer to the title question.

For the cases where QI resides in physical space, let us define
$v_{QI}$ by the extent of spacetime through which one quantum entity
can affect another. Consider the smallest cone which can bound this
region.
Let $\theta_{QI}$ be the angle of this cone from the vertical. Then
$v_{QI}=0$ corresponds to $\theta=0$, $v_{QI}=c$ to
$\theta_{QI}=\pi/4$ (coincides with the forward lightcone), and
$v_{QI}=\infty$  to $\theta_{QI}=\pi/2$.  But
these cases cover only half of spacetime, and one can in principle
have speeds ``faster than infinity'' for $\theta_{QI}> \pi/2$.  These
correspond to quantum entities which can interact with others
backwards through time.  For example, $\theta_{QI}=3\pi/4$ means that
QI can travel along the backward lightcone, which we define as
having `speed' $v_{QI}=$`$-c$'.  We can then conveniently
define an inverse speed,

\beq
w_{QI} = \cot\theta_{QI} = (v_{QI}/c)^{-1},
\eeq
so that the cases of $v_{QI}=c,\ \infty,$ and `$-c$' correspond to
$w_{QI}=1,\ 0,$ and $-1$.  For the rest of the paper, we will be
asking what it means for a quantum model to correspond to different
values of $w_{QI}$ (coupled with a characteristic spacetime).  We
summarize these results in Table I.  Categories are written such that
``Y'' is theoretically desirable (`spatial completeness' possibly
excepted), and we list ``Y'' when there is no reason the answer has to
be ``N''.  The final column is just opinion.

Let us dispense with the first line of Table I, which we will call a
`scripted universe'. Each point in space has all the QI it will ever
need from the start, and there is no transport of any (including
classical) information.  This universe of disjoint points then appears
to have dynamics in the same way that players appear to have an
impromptu conversation in a carefully written play.  This is
anathema to physics because dynamics are an illusion.  There are no
real causes and effects, just the script.

\begin{table*}
\caption{Regimes for the speed of QI.  The Y/N's answer
``Can there be a model in this regime with QI which is/has...'' E is
for ``effectively Y''. Strong causality is
violated in nonphysical space models without a preferred frame, though
weak causality is still preserved.  In $w_{QI}=-1$ models, strong and
weak causality are violated, though not beyond any $t$ cutoff.
The Y/N in the latter models is for finite/infinite space.}
\begin{tabular}{|c|c|c|l|l|c|c|c|c|c|c|c|}
\hline
&&&&     &&& $\tilde w$ Lorentz&preferred&one&spatially& \\
$w_{QI}$  & $\theta_{QI}$ & $v_{QI}$ & spacetime& example models &local? & causal? &  invariant? & frame? & time?& complete? &  nice?   \\
\hline
$\infty$& 0 &0&Disjoint&  Scripted&Y &n/a& Y & Y & Y& Y & N \\
\hline
1  & ${\pi\over4}$  & $c$&Local &Not allowed?  & Y & Y &Y&Y&Y&N&N?   \\
\hline
$(0,1)$ & $({\pi\over4},{\pi\over2})$&$(c,\infty)$& Not Local &Not allowed? \cite{Eberhard}& N&Y&N&N&Y&N&N\\
\hline
$\epsilon$ &${\pi\over2}-\epsilon$ & $c/\epsilon$ &Discrete& MWI, Bohm, Collapse& N&Y&E&N&Y&E&Y?\\
\hline
0  & ${\pi\over2}$  & $\infty$ & Continuous& MWI, Bohm, Collapse & N&Y&Y&N&N&Y&Y? \\
\hline
0  & ${\pi\over2}$  & $\infty$ & Nonphysical& MWI+space source? & N&Y/N$_s$&Y&N/Y&Y&Y&Y? \\
\hline
$-1$ & ${3\pi\over4}$ & `$-c$' & $t$ Cutoff& Unknown &N&Y&Y&N&Y/N&Y&N?\\
\hline
$-1$ & ${3\pi\over4}$ & `$-c$' & Acausal & Transactional\cite{transactional} &N&N$_{w,s}$&Y&Y&N&Y&N?\\
\hline
\end{tabular}
\end{table*}

The second possibility is  $v_{QI}=c$ ($w_{QI}=1$). Such a model
would be local and causal.  The question of Lorentz
invariance has two aspects, for one-way and two-way QI transport.
Recall that distance and time measurements are ultimately based upon
two-way communication of light, from the origin to a reference point
and back in a given frame.  We can then use such points to measure
one-way communication.  Thus let us define a normalized inverse speed
for the round trip to a reference point,

\beq
\tilde w = {{\rm QI\ round\ trip\ time} \over {\rm light\ round\ trip\ time}},
\eeq

\noindent
which can in principle be frame-dependent.  In whatever frame we use
to define $w_{QI}$ 
(which is the preferred frame if there is one), we have $\tilde w
=w_{QI}$ (for $w_{QI}\geq 0$).  For $v_{QI}=c$, we have $\tilde w
=w_{QI}=1$ in all frames, and thus two-way transport of QI in this
regime is Lorentz invariant.  But we still want to know if the one-way
transport picks out a preferred frame (apart from the one from
nonrelativistic quantum mechanics).  For the $v_{QI}=c$ case, there is
nothing {\it a priori} which does.

But is such a $v_{QI}=c$ model allowed?  As we said, experiments rule
out only $v_{QI}=c$ hidden variable models.  However, it also cannot
be a collapse model, because such collapses are inherently nonlocal.
Even the measurement of a single particle on a spacelike screen shows
this, since all the other points on the screen instantly know they can
no longer be the one to fire.

What about a noncollapse model such as the Many Worlds Interpretation
(MWI)?  By the MWI, we mean a quantum model based upon unitary
evolution of a universal wave function.  It would not entail new
`worlds' coming into existence with each quantum fork in the road
(which would be a horribly nonlocal phenomenon) but rather mundane
Schr\"{o}dinger evolution of a single wave function of the Universe
allowing arbitrary macroscopic superpositions.
QI in such a model could consist of that wave function, but possibly
also include information about a preferred basis for reality
\cite{pref basis}.  Can such a model be local?

No.  Consider {\it spatial completeness}, which means that some or all
dynamical `information'
describing a state of the Universe is present (`stored' or otherwise
accessible to dynamics there) at {\it every} point in space ${\bf x}$.
Spatially complete theories are necessarily nonlocal.  Classical
mechanics is spatially {\it in}complete because it is local: dynamical
information at $({\bf x_1},t)$ differs from that at $({\bf x_2},t)$.

For a quantum model to be local, it would need to be spatially {\it
in}complete, so that its QI differed from point to point.  But
dynamics in the MWI are encoded in a universal wave function, which is
spatially complete.  Thus the MWI is nonlocal.  There is only a single
wave function at time $t_0$, $|\Psi(t_0)\>$, with no provision for
different points having different $|\Psi(t_0)\>$. 
A spin singlet with particles at points ${\bf x_1}$ and
${\bf x_2}$ is represented by the wave function $[ |0;{\bf
x_1}\>|0;{\bf x_2}\> + |1;{\bf x_1}\>|1;{\bf x_2}\>]/\sqrt{2}$,
whether one speaks about it at ${\bf x_1}$, or ${\bf x_2}$, or any
other point.  It is a function of multiple spacetime points
and does not ``belong'' to any one of them.  It is this property which
makes the wave function nonseparable ({\it i.e.}, $\psi({\bf x_1},{\bf
x_2})$ cannot be written as $\phi({\bf x_1})\chi({\bf x_2})$), and
thus lies at the core of quantum nonlocality.  An augmented MWI could
have other QI such as preferred basis information or histories, which
could differ from point to point, but since the theory relies on
a spatially complete wave function, it is nonlocal.

An argument is sometimes made that the MWI is local because one cannot
measure EPR-like correlations until the measured subsystems are
brought together, {\it e.g.} that if Alice and Bob each measure half
of an EPR pair, they can see violations of Bell's inequalities only
after pooling their information in a relativistically causal way.  But
the wave function in the MWI, which contains QI about both
measurements, is present and exactly the same at both locations even
when they are spacelike separated.  That `information' is of course
classically unavailable, but we are interested in the properties of
QI.

There are other arguments that noncollapse models are local.
Deutsch and Hayden argue that the Heisenberg representation of the MWI
is local \cite{Deutsch}, but their formulation does not include
position operators.  When one does, position is again quantal and the
theory is again nonlocal.
Griffiths argues that there is no evidence for nonlocal influences in
the Consistent Histories Interpretation if one enforces a
`one-framework rule' \cite{Griffiths}.  However, the formalism of the
approach still contains a spacelike wave function, and thus has
nonlocal QI.

Quantum field theory does not seem to evade nonlocality either.  The
Lagrangian and canonical commutation relations respect locality, but
this ensures only that interactions and classical observables are
local.  The formalism of field theory still relies on a 
spatially complete quantum state.  Unless quantum theory can be
written using a formalism which is manifestly spatially {\it
in}complete, with QI at each point confined to its forward lightcone, one
cannot claim that the theory is truly local.

The third regime is $c < v_{QI} < \infty$ (distinguishably less than
$\infty$).  Such models would be spatially {\it in}complete and thus
could not include the MWI.  They would also be nonlocal and would
require a preferred reference frame in which $v_{QI}$ is
isotropic\cite{Eberhard}.  Further, even their two-way communication
breaks Lorentz invariance.  Viewed in a frame traveling at $\beta$
with respect to the preferred frame, $\tilde w' = \tilde w
(1-\beta^2)/(1-\beta^2 \tilde w^2)\simeq \tilde w (1-\beta^2)$ for
$\tilde w=w_{QI}$ close to zero.  Since $w_{QI}$ is distinguishable
from zero, we can tell that $\tilde w'$ differs from $\tilde w$.

Next come the discrete and continuous physical spacetimes with
infinite $v_{QI}$ (zero $w_{QI}$). They can accommodate collapse
models, 
or noncollapse models such as the Bohm model and the MWI.  They can be
spatially complete, are nonlocal, and require a preferred reference
frame with a corresponding preferred time.  We need to be careful
about defining the term `causal', though.  A weak criterion we would
want any causal theory to meet is that QI at each spacetime point
cannot affect QI in its past lightcone.  A stronger criterion would be
that there is at least one local frame in which QI cannot affect QI at
prior times. 
(`Relativistic causality' is even stronger, but identical
to locality.) Models with zero $w_{QI}$ satisfy weak causality, since
QI cannot propagate into the past lightcone, but strong causality is
satisfied as well, since in the preferred frame, no QI can affect an
earlier time.  One might argue that an infinite velocity, when viewed
in another frame, seems to go backward in time.  But we can simply
insist that the theory be defined in the preferred frame.  And as we
showed above, when $w_{QI}$ is immeasurably close to zero, $\tilde w$
appears the same in all frames.  Thus a model in these regimes can in
principle be written in a causal and Lorentz invariant way, except for
the imposition of a preferred frame.  We note that one can define a
set of preferred frames in a generally covariant way by specifying a
timelike unit vector $u^a$ at each point \cite{jacobson}.  The 
$u^a$ define spacelike hypersurfaces which one might say specify a
`preferred simultaneity' with each hypersurface having a unique
temporal index, $\tau$ (this requires that the hypersurfaces do not
cross each other, which is met if $u_{[a}\nabla_{b}u_{c]}=0$
\cite{ted}).  At each point on such a hypersurface, the
tangent vectors correspond to infinite velocity in the local preferred
frame.

In the discrete infinite case, $v_{QI} =c/\epsilon$ with $\epsilon =
w_{QI}$ fixed but indistinguishable from zero.  This follows if QI can
travel the length of the Universe in a single unit of time, {\it i.e.},
$\epsilon < c \Delta \tau /L_{Universe}$ \cite{Note discrete}.
For example, if $\Delta \tau\sim t_{Planck}\sim 10^{-43}$ sec, and
$L_{Universe}\sim 10^{10}$ lyr, then if $v_{QI}>10^{61}c$\cite{Gisin},
one can never distinguish it from $\infty$.  Such models can be
effectively spatially complete, since QI at each location is
at most $\Delta \tau$ out of date.  All QI dynamics could happen in
sub-$\Delta \tau$ time, according to some new parameter $\tqi$, which
cycles through those dynamics each $\Delta
\tau$.  In some sense this is a second `time', which governs QI
dynamics.  But in this regime we can also think of $\tau$ and $\tqi$
as different aspects of a single underlying time $T$ via a mapping
like

\beq
\tau=\Delta \tau \ {\rm int}(T/\Delta \tau),\ \tqi= T\, {\rm mod} \Delta \tau.
\eeq

In the continuous case, $\Delta \tau \rightarrow 0$ 
and $v_{QI}=\infty$.  Many physicists, if pressed, would probably pick
this regime since it seems that the wave function is updated
everywhere ``instantly'', {\it i.e.}, it is spatially complete.  A
problem with this regime is that, since all QI dynamics and transport
take place in zero $\tau$, one really needs a second `time' because
the above mapping becomes singular.
One cannot base these dynamics (whatever they are) on $\tau$,
which along each preferred hypersurface is constant.  But what does it
mean to have two times? Picture a curve $\gamma(T)$ tracing a path
through a cylindrical $\tau-\tqi$ space.  The path is a helix going
through all values of $\tqi$ before advancing infinitesimally in $\tau$.
Thus the helix has vanishing pitch.
In that limit, the curve $\gamma(T)$ becomes two-dimensional, because
any point belonging to the two-dimensional $(\tqi,\tau)$ space ({\it
e.g.}
${\mathcal{S}}^1\times {\mathcal{R}}$) also belongs to the curve
$\gamma(T)$ (this is analogous to a singular mapping of a line to a
plane, which are equivalent infinite sets).
Thus in this regime we need two different temporal parameters: one for
the observed temporal evolution to each new hypersurface, and one for
QI transport dynamics along each hypersurface.  But unless one can
probe QI transport dynamics, one cannot detect the preferred frame or
the $\tqi$ dimension, and all observations will correspond to the
usual SO(3,1) Lorentz invariance with one apparent time.

All the above regimes are specified by QI and the spacetime on which
it exists.  Now suppose that QI does not exist on spacetime, but
rather spacetime information, such as the metric, is part of QI.  Then
space would be `nonphysical' in the sense that QI would neither be
stored in nor transported through space.  QI would exist elsewhere,
such as a reified Hilbert or Fock space.  If QI encodes all spacetime
behavior, space has at most a secondary role.  This is appealing from
the quantum point of view because the spatial wave function
$|\psi({\bf x}_1, {\bf x}_2, t)\> =\< {\bf x}_1|\<{\bf x}_2|\psi(t)\>$
is only one possible representation of  $|\psi(t)\>$.  In
such a model, all QI could be instantly associated with any spatial
point (though without actually needing transport there), which is like
the infinite speed cases above.  So we assign this case $w_{QI}=\tilde
w=0$.
(If local quantum models were allowed, one could construct a
nonphysical space version, which would have the same properties as
$w_{QI}=1$ physical space models.)
Like those cases,
this case is nonlocal and can be spatially complete (now even
spacetime information could be spatially complete). For causality,
there are two possibilities.  If the theory has a preferred frame,
then, just as in physical space, weak and strong causality are
satisfied.  But it is possible that the theory is completely
Lorentz-invariant.  In that case, the theory
would be only weakly causal because it is the preferred frame which
allows us to specify a unique past and future for spacelike
interactions.  In either case, we do not need the second time $\tqi$
since there is no need to transport QI.

What is the role of `space' here?  First, it could be just an
abstract entity, a convenient way of expressing certain properties of
QI.  Then it should be possible to write the theory without reference
to space at all.  It is not clear how to do this, especially
when one brings in gravity. Second, space could be a construct arising
from QI.  Then space would still exist in some sense and could have a
dynamical role, but it would ultimately be explained solely in terms
of QI.  We note that recently there have been attempts to construct
discrete (extra) spatial dimensions out of gauge degrees of freedom
\cite{ACG}, but these
do not yet include a complete description of gravity.

The fundamental question for nonphysical space models is not ``why is
there nonlocality'', 
since nonlocality can arise trivially,
but ``why is there locality?''  Why do classical observables
have a limiting speed $c$ if QI (which must account for all classical
information) is unfettered by space at all?  How would such a theory
meld with general relativity where space itself has a dynamical role?
Some argue
\cite{Rovelli} that general relativity prefers space to be
nonphysical, with all (quantum) dynamics occurring via the relations
between entities.  But this might make things more complicated.  For
example, to specify $N$ points purely relationally requires of order
$N^2$ numbers instead of order $N$ (where $N$ is huge).

Finally we have the regimes with ``$v_{QI}> \infty$'', where QI can
travel into the past.  If any QI can travel arbitrarily far backwards
in time, then $w_{QI}$ is really $-\infty$.  However, we refer to this
case as $w_{QI}=-1$ because the only Lorentz-invariant choice is to
restrict propagation to the forward and backward 
lightcones \cite{transactional}.  Here the
definition of ``round trip'' is ambiguous.  It can mean return as
close as possible to the starting spacetime point, in which case
$\tilde w=0$ (zero round trip time), or as ``fast'' a return as
possible, in which case we could have $\tilde w=-1$ (negative round
trip time).  Both of these possibilities are Lorentz invariant.
While classical observables seem causal, the
underlying QI dynamics would depend upon QI from the distant past and
future.  We again need a second time to parametrize this, but it is
less clear what this means since causality for QI is, in a real sense,
lost.

Instead, one could have $w_{QI}=-1$ with a hard
cutoff on how far QI can propagate backwards in time.  
Such a model would be causal in macroscopic time.  In fact, at scales
above the cutoff, this case is identical to the infinite $v_{QI}$
cases with the cutoff picking out the preferred frame.  
Although we again need both $\tau$ and $\tqi$, if 
classical evolution happens only on scales above the cutoff,
and space is finite, then backwards and forwards
evolution of $\tau$ in $\tau$--$\tqi$ space can in principle be
represented with a nonsingular mapping $\gamma(T)$,  with 
one underlying time $T$.  But if space is infinite, the mapping becomes
singular.

So it is useful to characterize quantum models by $v_{QI}$.  Since the
MWI is spatially complete, it cannot have $v_{QI}=c$ and thus is
nonlocal.  It is probably not possible to have any local dynamical
quantum model, in which case the appealing regimes are those with
infinite effective speed.  Continuous physical space is appealing from
the classical point of view, but such models require a second time for
QI transport.  Discrete models with $v_{QI}=c/\epsilon$ avoid this,
but only if the Universe is finite \cite{Note discrete}.  Nonphysical
space, where spacetime information is part of or arises from QI, is
appealing from the quantum point of view, but no complete model exists
yet.  This tension between classical local spacetime and quantum
nonlocality will need to be resolved in any quantum theory of gravity.

We thank Carlo Rovelli, Nicholas Gisin, Ted Jacobson, Rashmi Ray, Bob
Griffiths, and Seth Lloyd for helpful comments.


\begin{thebibliography}{99}
\bibitem{Note discrete} If the Universe is infinite, either now or through 
infinite-time expansion, then $v_{QI}$ needs to be truly infinite for quantum 
mechanics never to fail.
\end{thebibliography}
\end{document}